\documentclass[journal=langd5,manuscript=article,layout=traditional]{achemso}

\usepackage[T1]{fontenc}
\usepackage[english]{babel}
\usepackage{amsmath}
\usepackage{graphicx}
\usepackage{subfig}
\usepackage[usenames, dvipsnames]{color}
\usepackage{chemformula}
\usepackage{upgreek}
\usepackage[breaklinks,colorlinks,citecolor=blue,linkcolor=blue,urlcolor=black]{hyperref}
\setcitestyle{numbers,square,sort&compress}
\setkeys{acs}{maxauthors=2, etalmode=truncate}
\SectionNumbersOn

\author{Michael Ludwig}
\author{Regine von Klitzing}
\email{klitzing@smi.tu-darmstadt.de}
\affiliation[TUD]
{Soft Matter at Interfaces, Department of Physics, Technische Universit\"at Darmstadt, Hochschulstrasse 8, D-64289 Darmstadt, Germany}

\title{Untangling superposed double layer and structural forces across confined nanoparticle suspensions}

\setkeys{acs}{keywords = true}

\begin{document}

\begin{abstract}

The description of forces across confined complex fluids still holds many challenges due to the possible overlap of different contributions. Here, an attempt is made to untangle the interaction between charged surfaces across nanoparticle suspensions. Interaction forces are measured using colloidal-probe atomic force microscopy. The experimental force profiles are considered as a superposition of double layer and structural forces. In order to independently describe the decay of the double layer force, the ionic strength of the suspension is determined by electrolytic conductivity measurements. Jellium approximation is used to define the impact of the fluid on screening the surface potential. There, the nanoparticles are considered homogeneously distributed across the fluid and screening is only carried out via the particles counterions and added salt. The structural force follows a damped oscillatory profile due to the layer-wise expulsion of the nanoparticles upon approach of both surfaces. The description of the oscillatory structural force is extended by a depletion layer next to the confining surfaces, with no nanoparticles present. The thickness of the depletion layer is related to the electrostatic repulsion of the charged nanoparticles from the like-charged surfaces. The results show that the total force profile is a superposition of independent force contributions without any mutual effects. Using this rather simple model describes the complete experimentally determined interaction force profiles very well from surface separations of a few hundred nanometres down to the surfaces being almost in contact.

\end{abstract}

\section{Introduction}

Interactions between two surfaces across a liquid medium are of fundamental interest in colloid and interface science. They determine the (macroscopic) behaviour in a wide range of systems, such as dispersions, emulsions, lipid bilayers or foams. Exact knowledge of the interactions is interesting for basic research, as the prediction of the phase behaviours of colloidal systems showed \cite{vanRoij.1999,Dijkstra.2000}. Moreover, it is crucial in various industrial applications, such as in food industry \cite{Dickinson.2015}.

Interactions are typically described by the classical theory of Derjaguin, Landau, Verwey, and Overbeek, the DLVO-theory \cite{Derjaguin.1941,Verwey.1947}. The DLVO-theory describes the interactions as a superposition of van der Waals (dispersion) and diffuse double layer (electrostatic) interactions. This theory has been proven to be extremely reliable, especially in diluted salt systems. The double layer force is typically described by the Debye-H\"uckel (DH) approximation \cite{Russel.1989, Evans.1999, Israelachvili.2011, Butt.2018}. It may, however, deviate from DH approximation in various cases. Charge-regulation effects were found to be important at small surface separations \cite{Trefalt.2016b} and ion pairing in symmeteric multivalent electrolyte solutions results in larger screening lengths than obtained from DH-approximation \cite{Smith.2019b}. An ongoing debate concerns the underscreening in concentrated electrolytes \cite{Wang.2013,Smith.2016} and ionic liquids \cite{Gebbie.2017}, where the range of the double layer force even increases above a certain electrolyte concentration, a fundamental contradiction to the DH approximation.

Different types of forces outside the DLVO framework were discovered and summarized as non-DLVO forces in more complex fluids \cite{Christenson.1988}. When surfaces interact in dispersions containing non-adsorbing objects for example, the well-known depletion attraction can be observed \cite{Asakura.1954,Lekkerkerker.2011}. At higher volume fractions of the dispersed phase, oscillatory depletion forces can emerge. Oscillatory depletion forces in dispersions containing polyelectrolytes,\cite{Kleinschmidt.2009,Biggs.2010,Uzum.2011} micelles \cite{Richetti.1992,Basheva.2007,Christov.2010,Tabor.2011,Anachkov.2012}, or solid nanoparticles \cite{Piech.2002,Tulpar.2006,Zeng.2011} have been determined experimentally, using the surface forces apparatus (SFA), colloidal-probe atomic force microscopy (CP-AFM), or a thin film pressure balance (TFPB). Moreover, theoretical studies of oscillatory depletion forces were carried out for dispersions of hard-spheres \cite{Kralchevsky.1995,Roth.2000,Trokhymchuk.2015} and for dispersions of charged objects \cite{Chu.1995,Klapp.2008b}. The interplay of bulk structure and oscillatory depletion force was emphasised by a comparison of direct force measurements and bulk scattering techniques, such as small-angle X-ray scattering (SAXS) \cite{Klapp.2007,Klapp.2008, Tabor.2011b}. Since the period of the force oscillations is determined by the bulk pair correlation function of the dispersed objects - and, therefore, their structuring in the bulk fluid - the term oscillatory structural force was introduced and will be used throughout this study.

While a lot of work was spent on the description of oscillatory structural forces itself, \textit{e.g.} by the authors of this paper, only a few research dealt with the role and superposition of DLVO-type forces - mainly the double layer forces - in such fluids. A first description of the interplay between double layer and depletion attraction was obtained for surfaces of two oil-droplets in a micellar solution \cite{MondainMonval.1995}. Due to the presence of a depletion attraction, the interaction force profile deviates from a exponential decay, as typically observed for a pure double layer interaction. Recently, the interaction force profiles between charged silica surfaces in dispersions containing like-charged polyelectrolytes (polystyrene sulphonate, PSS) were modelled \cite{MoazzamiGudarzi.2016,MoazzamiGudarzi.2017}. There, the polyelectrolytes are treated as multivalent ions and non-linearised PB solution is assumed for the description of the short-ranged double layer force \cite{Tellez.2011}. At larger surface separations oscillatory structural forces dominate the interaction force profile.

Treating the dispersed objects as multivalent ions results in unexpectedly large screening leaving it a heavily discussed topic. Moreover, the complete depletion of the dispersed objects at small distances between the confining surfaces is still missing in the existing models. 

The aim of this work is to untangle the superposed force contributions. Therefore, interaction forces between charged silica surfaces in suspensions containing nanoparticles and salt at various concentrations are measured. Additionally, the screening behaviour of suspensions is determined independently from electrolytic conductivity measurements. In that way, the double layer force could be extracted from the oscillatory structural force, leading to a deeper insight in the interaction of surfaces across complex fluids.

\section{Experimental section}

\subsection{Materials}

Colloidal silica nanoparticle (NP) suspensions (Ludox HS-30) were purchased from Sigma Aldrich (Germany). Ludox is a registered trademark of W.R. Grace \& Co.-Conn. for discrete, spherical NPs of amorphous silica. The NPs are dispersed in water, with sodium as counterions of the negatively charged surface groups. The NP diameter and distribution was determined by transmission electron microscopy (FEI CM20 microscope, The Netherlands) as $15.8\,\pm\,2.9\,\text{nm}$. The density of the suspension was determined by weighing suspensions with known concentrations using a high precision volumetric flask. Setting the density of water to $0.998\,\text{g}\,\text{cm}^{-1}\ (20\,^\circ\text{C})$ the density for silica was determined as $1.97\,\pm\,0.07\,\text{g}\,\text{cm}^{-1}$. The original stock of NP suspension was dialysed in dialysis tubes (SnakeSkin, 3.5k MWCO, ThermoFisher, Germany) with deionised water (milliQ-grade, $18.2\,\text{M}\Omega\cdot\text{cm}$ resistivity, Merck, Germany) for ten days, with daily water exchange. Deionized water was used throughout the whole project. This dialysis method (diffusion dialyis) is able to extract certain impurities, yet a considerable amount of counterions remains in solution \cite{Iler.1979}. After dialysis, the NP concentration was characterized by weighing the sample before and after drying ($24\,\text{h}$ at $80\,^\circ\text{C}$ and vacuum). The dialysed stock suspension was then diluted with aqueous sodium chloride (NaCl, Suprapure 99.99, Merck, Germany) solutions to achieve the desired concentrations assuming ideal mixing.

\subsection{Methods}

\subsubsection{Conductivity measurements}

The conductivities of the suspensions were measured using a Nano Zetasizer (Malvern Instruments, UK). Before each measurement, the measuring cell was rinsed thoroughly with water and absolute ethanol and dried in a nitrogen stream. One data point is averaged by at least five measurements to ensure reproducibility.

\subsubsection{Substrates and Colloidal Probes} 

Silicon wafers (Soitec, France) were used as substrates. Prior to the experiments, the wafers have been cleaned by the Piranha method, using a 1:1 (vol/vol) mixture of hydrogen peroxide ($30\,\%$, Th. Geyer, Germany) and sulphuric acid ($96\,\%$, Carl Roth, Germany). With this method, a very smooth native silica layer is formed on the silicon surface. Afterwards, the wafers were rinsed with large amounts of water and dried in a nitrogen stream. Non-porous silica particles (Bangs Laboratories, USA) with a diameter of $5\,\upmu\text{m}$ were used as colloidal probes. One particle was glued (UHU Endfest Plus 300, UHU, Germany) to the end of a tipless rectangular cantilever (SD-qp-SCONT-TL, Nanosensors, Switzerland) using a three-dimensional microtranslation stage. Immediately before the experiment, both, cantilever and substrate, were cleaned by exposure to oxygen plasma (Diener Femto, Germany) to remove all organic impurities.

\subsubsection{Force measurements}

Force measurements were carried out using a Cypher ES atomic force microscope (Asylum Research, USA). Before each experiment, the spring constants of the cantilevers were determined as $0.018 - 0.026\, \text{N}\,\text{m}^{-1}$ using the method described by Sader \textit{et al.} \cite{Sader.1999}.
After calibration, cantilever and substrate were completely immersed within the NP suspensions. The temperature was set to $20.0\,^\circ\,\text{C}$ via a cooler-heater sample stage and left to equilibrate for at least 30 minutes. The cantilever deflection was measured as a function of the position of the z-piezo. Using Hooke's law, the cantilever deflection was converted into force. To convert the raw data into force $F$ $vs.$ separation $h$ curves (in the following described as force profiles), well-known algorithms are used \cite{Ducker.1991, Butt.2005}. The Derjaguin approximation was used to normalise the measured force $F$ against the effective radius $R_{\text{eff}}$ \cite{Evans.1999, Israelachvili.2011}. For a planar plate $vs.$ sphere, $R_{\text{eff}}$ is simply the colloidal probe radius $R$. The resulting force is, therefore, in the following described as $\frac{F}{R}$.

The starting point of each measurement was set to $1000\,\text{nm}$ with an approach-retraction velocity of $100\,\text{nm}\,\text{s}^{-1}$ and a sampling rate of $2\,\text{kHz}$. The surfaces were assumed to be in contact once the normalised force exceeds $2.5\,\text{mN}\,\text{m}^{-1}$ (constant compliance region). For each system, at least 30 individual force profile were averaged to ensure reproducibility and to substantially increase the force resolution to less than $0.002\,\text{mN}\,\text{m}^{-1}$.

\subsection{Fitting of interaction forces} \label{sec:fitting}

The normalised interaction forces $\frac{F}{R}$ between two charged surfaces at separations $h$ were fitted using a superposition of double layer $F_{\text{dl}}$ and structural forces $F_{\text{str}}$. 

\begin{equation}
\frac{F}{R}\left(h\right) \ = \ \frac{F_{\text{dl}}}{R}\left(h\right) + \frac{F_{\text{str}}}{R}\left(h\right)
\end{equation}

The Hamacker constant of the type of silica colloidal probe used was previously determined as $0.07\,\times\,10^{-21}\,\text{J}$ \cite{Valmacco.2016}. Van der Waals forces are, therefore, neglected in the following analysis, since they are small and short ranged under the studied conditions.

\subsubsection{Double layer forces}

The overlap of electric double layers of two surfaces induce the double layer force. The effective potentials $\psi_{\text{eff}}$ of both confining silica surfaces are assumed to be equal, giving rise to a symmetric potential. Linearised Poisson-Boltzmann theory (also known as Debye-H\"uckel (DH) approximation) was used to model the normalised double layer force $\frac{F_{dl}}{R}$ in dependence of the surface separation $h$ by the equation \cite{Evans.1999, Israelachvili.2011}:

\begin{equation} \label{equ:double_layer_force}
\frac{F_{\text{dl}}}{R}\left(h\right) \ = \ 4 \pi \epsilon \epsilon_{0} \kappa \psi_{\text{eff}}^{2} \cdot \exp\left(-\kappa h\right)
\end{equation}

where $\epsilon$ is the dielectric constant of water, $\epsilon_{0}$ is the dielectric permittivity of vacuum. 

The Debye screening length $\kappa^{-1}$ is defined as following:

\begin{equation} \label{equ:debye_screening}
\kappa^{2} \ = \ \frac{2e^{2} N_{\text{A}} I}{\epsilon \epsilon_{0} k T}
\end{equation}

where $e$ is the elementary charge, $N_{\text{A}}$ is the Avogadro number, $k$ is the Boltzmann constant and $T$ is the temperature. 

\begin{figure}[ht]
\begin{center}
\includegraphics[width=0.4\textwidth]{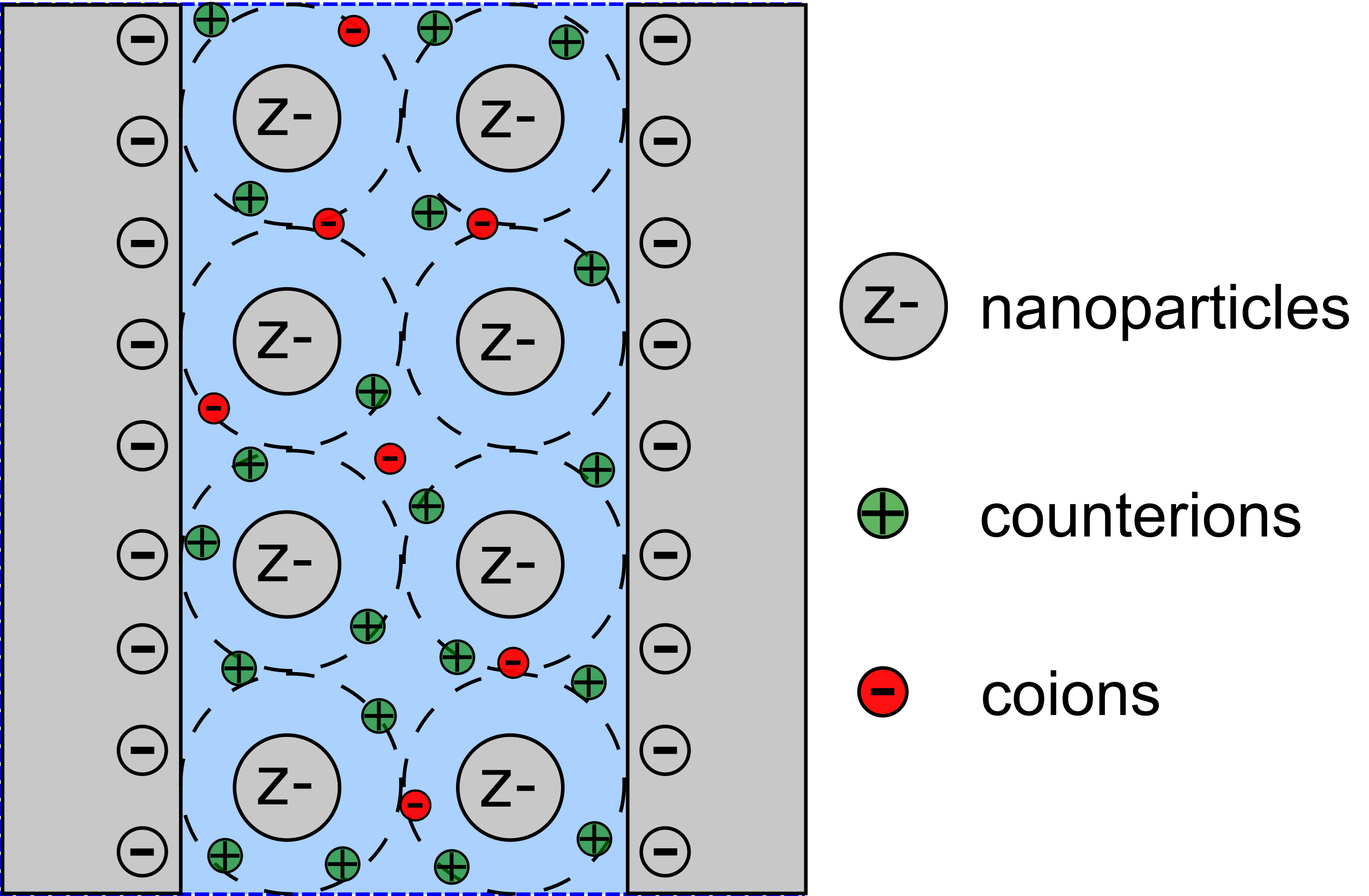}
\caption{Schematic representation of the jellium approximation for dispersions containing charged NPs confined between charged walls. The NPs are assumed to be uniformly distributed across the fluid due to their repulsive interaction, while monovalent ions (the particles' counterions and added salt) obey Boltzmann distribution due to the surface potential. As a result, the NPs \textit{do not} contribute to Debye screening of the surfaces' electric field.}
\label{fig:jellium_approx}
\end{center}
\end{figure}

In this study the ionic strengths $I$ of the suspension is calculated using the jellium approximation (JA). A schematic representation of the JA is given in figure\ref{fig:jellium_approx}.

Historically, the term 'jellium' was introduced for quantum mechanical description of quasi-free electrons in a solid where the positive atomic nuclei are assumed to be uniformly distributed in space \cite{Giuliani.2005}. This framework was later adopted for the description of colloidal systems at low concentrations of electrolyte \cite{BeresfordSmith.1983,BeresfordSmith.1985}. The dispersed NPs are considered as a scaffold with the monovaltent ions distributed around. The concentration of NPs $c_{p}$ is assumed constant. Monovalent cations $c_{+}$ and anions $c_{-}$ (counterions of the particles and added salt ions), however, obtain Boltzmann distribution. Their concentration next to a charged surface, with an effective surface potential $\psi_{\text{eff}}$, can be calculated using the ions valence $z_{i}$ (+1 for counterions, -1 for coions).

\begin{equation}
\begin{aligned}
c_{p} = c_{p,\infty} &= \text{const.}, \\
c_{+} = c_{+,\infty} \exp{\left(\frac{- z_i e \psi_{\text{eff}}}{k T}\right)}&, \ \ c_{-} = c_{-,\infty} \exp{\left(\frac{- z_i e \psi_{\text{eff}}}{k T}\right)}
\end{aligned}
\end{equation}

As a result, using the jellium approximation, the NPs are neglected from Debye screening. The ionic strength $I$ is only depending on the concentrations of the monovalent ions $c_{i}$.

\begin{equation}
I \ = \ \frac{1}{2} \sum \limits_{i} z_{i}^{2} c_{i}
\end{equation}

Being able to calculate the ionic strength $I$ of the suspension leaves the effective potential $\psi_{\text{eff}}$ as the only free parameter that enters the fit to the double layer force.

At higher surface potentials, the fitted effective surface potential $\psi_{\text{eff}}$ is known to deviate from the diffuse layer potential $\psi_{\text{dl}}$, \textit{i.e.} the real potential at the origin of the diffuse layer \cite{Smith.2019}. The effective potential can be converted into the diffuse layer potential $\psi_{\text{dl}}$ assuming a symmetric 1:1 electrolyte ($z$\,=\,1) \cite{Evans.1999}.

\begin{equation} \label{equ:diffuse_layer_pot}
\psi_{\text{eff}} = \frac{4 k T}{z e} \tanh{\left(\frac{z e \psi_{\text{dl}}}{4 k T}\right)}
\end{equation}

The surface charge density $\sigma$ can then be calculated from the diffuse layer potential $\psi_{\text{eff}}$ assuming that the total charge of the double layer must neutralize the surface charge (electroneutrality). This is also known as the Grahame equation \cite{Evans.1999, Israelachvili.2011}.

\begin{equation} \label{equ:Grahame}
\sigma = \frac{2 k T \epsilon \epsilon_{0} \kappa}{z e} \sinh{\left(\frac{z e \psi_{\text{dl}}}{2 k T}\right)}
\end{equation}

The applicability of the JA  which will be further considered in the discussion chapter. In general discussions, the term 'macroion' is used to describe the NPs, since this model can be extended to other colloidal systems, \textit{e.g} to dispersions containing ionic micelles or polyelectrolytes. 

\subsubsection{Structural forces}

After subtraction of the fitted double layer force from the experimental data, the structural force was fitted to equation \ref{equ:structural_force}. 

\begin{equation} \label{equ:structural_force}
\frac{F_{\text{str}}}{R}\left(h\right) =
\begin{cases}
\frac{F_{\text{str}}}{R}\left(h'\right) \ = \ \text{const.}, \ \ & 0 < h \leq h' \\
- A \cdot \exp\bigl(\frac{-h}{\xi}\bigr) \cdot \cos\bigl(\frac{2 \pi}{\lambda} (h - h') \bigr), \ \ & h' < h < \infty \\
\end{cases}
\end{equation}

\begin{figure}[ht]
\begin{center}
\includegraphics[width=0.4\textwidth]{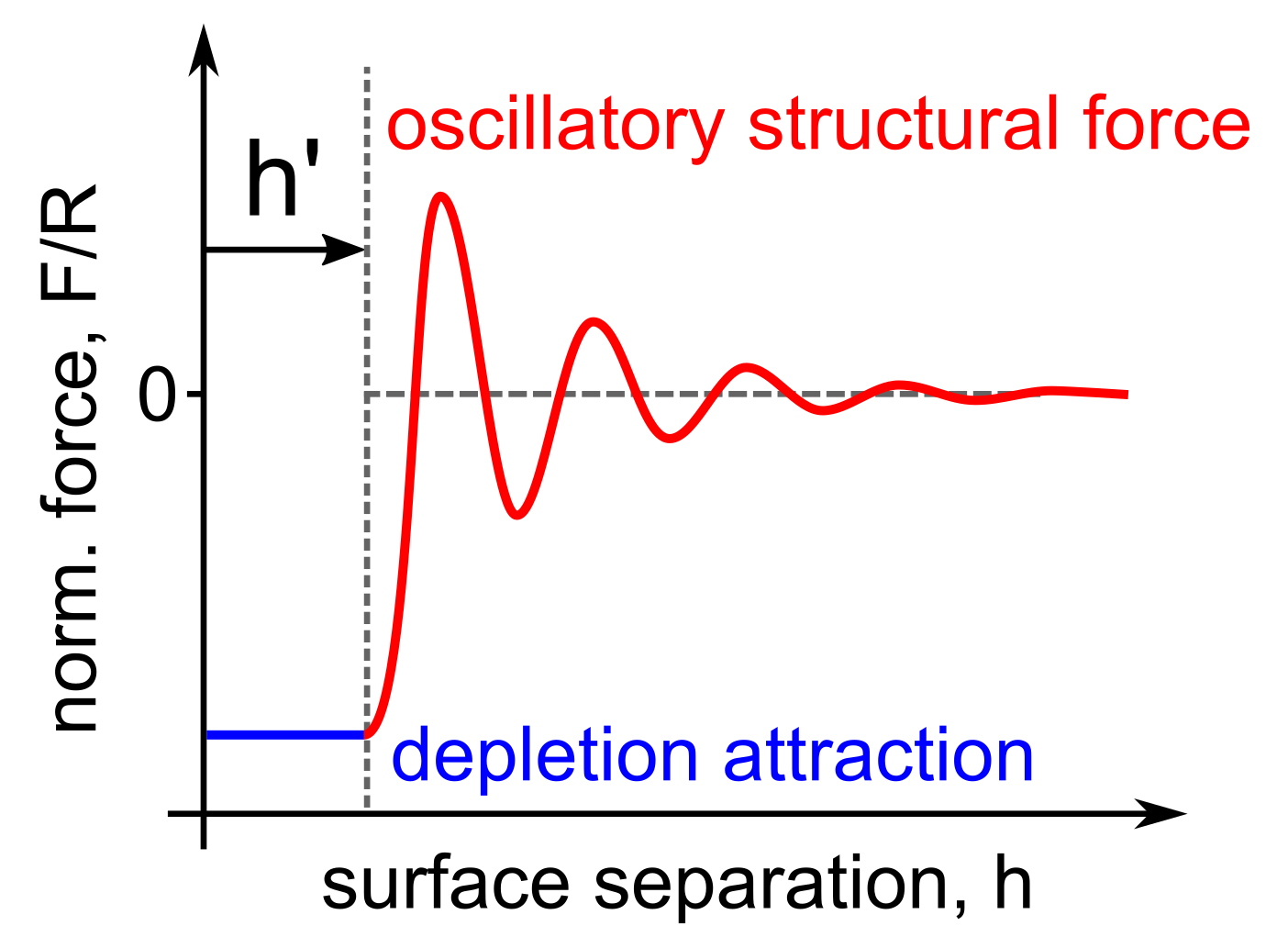}
\caption{Schematic model of the total structural force. Below a certain offset $h'$ the interaction switches from an oscillatory structural force to a depletion attraction.}
\label{fig:depletion_interaction}
\end{center}
\end{figure}

Combination of the oscillatory structural force ($h' < h < \infty$) with the depletion attraction ($0 < h \leq h'$) is referred to as the structural force. The total structural force is schematically illustrated in figure \ref{fig:depletion_interaction}.

First, the asymptotic behaviour was fitted using a damped oscillatory profile. This contribution is referred to as the oscillatory structural force, with the amplitude $A$, the decay length $\xi$, and the wavelength $\lambda$. The parameter $h'$ corrects for the offset of the structural force. The oscillatory structural force is used to describe the structural force for surface separations $h$ being larger than the offset $h'$.

At surface separations $h$ smaller than the offset $h'$, all nanoparticles are considered to be depleted from the vicinity of the surfaces, resulting in a constant depletion attraction. The depletion attraction is typically calculated from the bulk osmotic pressure. Here, the strength of the depletion attraction is set as a constant value. It is determined from the value of the fitted oscillatory structural force at the offset $h'$, \textit{i.e.} the first minimum of the oscillatory structural force, to ensure continuity between both scenarios. 

It was previously shown, that the fit parameters of the oscillatory structural force may vary, \textit{i.e.} the fit parameters are not independent of the region the oscillatory structural force is fitted to \cite{Schon.2018}. The values of the fit parameters typically vary with the same period of the force profile for different starting points of the fit. To account for this problem, the starting point of the fit is varied over at least the first period. Exception: For the suspension of a NP volume fraction of $\phi$\,=\,0.031 and addition of 5\,mM NaCl the starting point was only varied half an oscillation after the first minimum in the force profile, since no consistent fit was obtained above, due to a weak structural force. Error bars refer to the standard deviations from all individual fits with different starting points in the force profile.

\section{Results}

The ionic strength of each suspension used in this study is calculated from its electrolytic conductivity (section \ref{sec:ionic_strength}). Individual determination of the ionic strength eliminates it as free parameter in fitting of the double layer force.

Direct force measurements were carried out between silica surfaces in suspensions containing different amounts of nanoparticles (NPs) and monovalent salt (section \ref{sec:exp_afm}). Interaction profiles are described as a superposition of double layer forces from the outer confining surfaces and structural forces induced by the confined NP suspensions.

\subsection{Determination of the ionic strength of nanoparticle suspensions} \label{sec:ionic_strength}

\begin{figure}[ht]
\begin{center}
\includegraphics[width=0.4\textwidth]{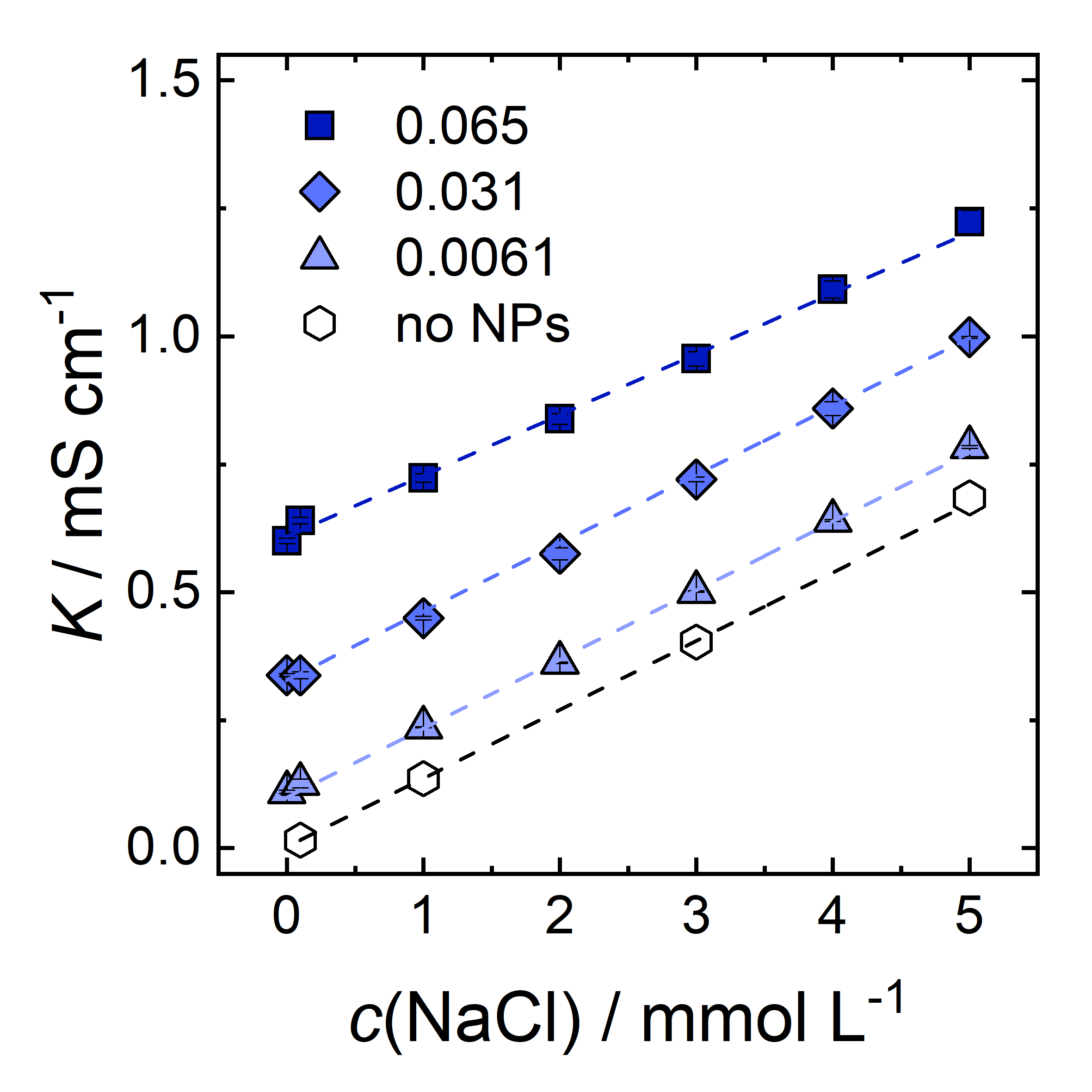}
\caption{Electrolytic conductivities $K$ of suspensions at varying volume concentrations of dispersed NPs as a function of added NaCl. Dotted lines are linear fits to the data.}
\label{fig:conductivity}
\end{center}
\end{figure}

In order to obtain the ionic strength of pure NP suspensions, their electrolytic conductivities were measured. The results for pure suspensions and suspensions with added NaCl are displayed in figure \ref{fig:conductivity}. In general, contributions to the conductivity are expected from protons \ch{H+} and hydroxide-anions \ch{OH-}, as well as from the negatively charged silica nanoparticles NP$^{\text{Z}-}$ including their sodium counterions \ch{Na+}.

The pH value of the suspension was determined as 9.2 (figure S1, ESI), revealing the concentrations of protons \ch{H+} and hydroxide-anions \ch{OH-} to be rather low. Their contribution is neglected in the following. The mobility of the NPs is also small compared to the one of its \ch{Na+} counterions. Consequently, the ionic strength of the pure silica suspensions is dominated by the NPs counterions. Ion-ion correlations, as described by the Kohlrausch law \cite{Atkins.2018}, are neglected. The concentration of \ch{Na+} counterions $c_{0}$(\ch{Na+}) in suspensions without addition of salt is, therefore, calculated by the limiting molar conductivity of the sodium counterions ($\lambda_{+}^{0}$(\ch{Na+}) = $5.011$\,mS\,m$^2$\,mol$^{-1}$) and the electrolytic conductivity without added salt $K_0$. Determined ionic strengths of the pure NP suspensions $I_{0}$ are summarized in Table \ref{tab:fit_conductivity}.

\begin{equation} \label{equ:ionic_strength}
c_{0}(\ch{Na+}) = \frac{K_0}{\lambda_{+}^{0}(\ch{Na+})}, \ \ \ I_{0} \ = \ \frac{1}{2} \, c_{0}(\ch{Na+})
\end{equation}

\begin{table}[ht]
\centering
\begin{tabular}{ c c c c }
\hline
$\phi$\,(\ch{SiO2}) & $b$ & $K_0$ & $I_{0}$ \\
 &(mS\,L\,mmol$^{-1}$\,cm$^{-1}$)&(mS\,cm$^{-1}$)&(mmol\,L$^{-1}$)\\
\hline
- & \ $0.134\,\pm\,0.001$ \ & \ $0.002\,\pm\,0.001$ \ & $0.02\,\pm\,0.01$ \\
$0.0061$ & $0.135\,\pm\,0.002$ & $0.099\,\pm\,0.004$ & $0.99\,\pm\,0.04$ \\
$0.031$ & $0.135\,\pm\,0.001$ & $0.329\,\pm\,0.006$ & $3.28\,\pm\,0.06$ \\
$0.065$ & $0.119\,\pm\,0.005$ & $0.609\,\pm\,0.006$ & $6.08\,\pm\,0.06$ \\
\hline
\end{tabular}
\caption{Parameters from the linear fit ($K$\,=\,$b$\,$c$(NaCl)\,+\,$K_0$) on the conductivity data shown in figure \ref{fig:conductivity}. The value for the ionic strengths of the pure NP suspensions $I_{0}$ without added salt are calculated via equation \ref{equ:ionic_strength}.}
\label{tab:fit_conductivity}
\end{table}

The slopes $b$ of all fitted curves are approximately the same, although a small deviation for the highest NP concentration is observed. Accounting for this deviation would decrease the calculated ionic strength in the system by less than 5\,\%.  Added NaCl (within the measured concentrations) is, therefore, assumed to be completely dissociated and not to be adsorbed onto the NPs. The total ionic strength $I$ can simply be calculated as the sum of the ionic strength of the pure NP suspension $I_{0}$ and the ionic strength from the added salt $I_{\text{NaCl}}$ under the assumption of full dissociation.

\begin{equation} \label{equ:total_ionic_strength}
I = I_{0} + I_{\text{NaCl}}
\end{equation}

\subsection{Interaction forces} \label{sec:exp_afm}

In order to highlight different characteristics of the interaction forces, two representations (semilogarithmic and linear) for the same force profiles are chosen in this section. First, a semilogarithmic representation is preferred to analyse the double layer contribution (section \ref{subsubsec:exp_dl}). The second part focusses on the structural force, and, therefore, a linear representation is selected (section \ref{subsubsec:exp_structural}).

\subsubsection{Double layer force} \label{subsubsec:exp_dl}

\begin{figure}[p]
\begin{center}
\includegraphics[width=\linewidth]{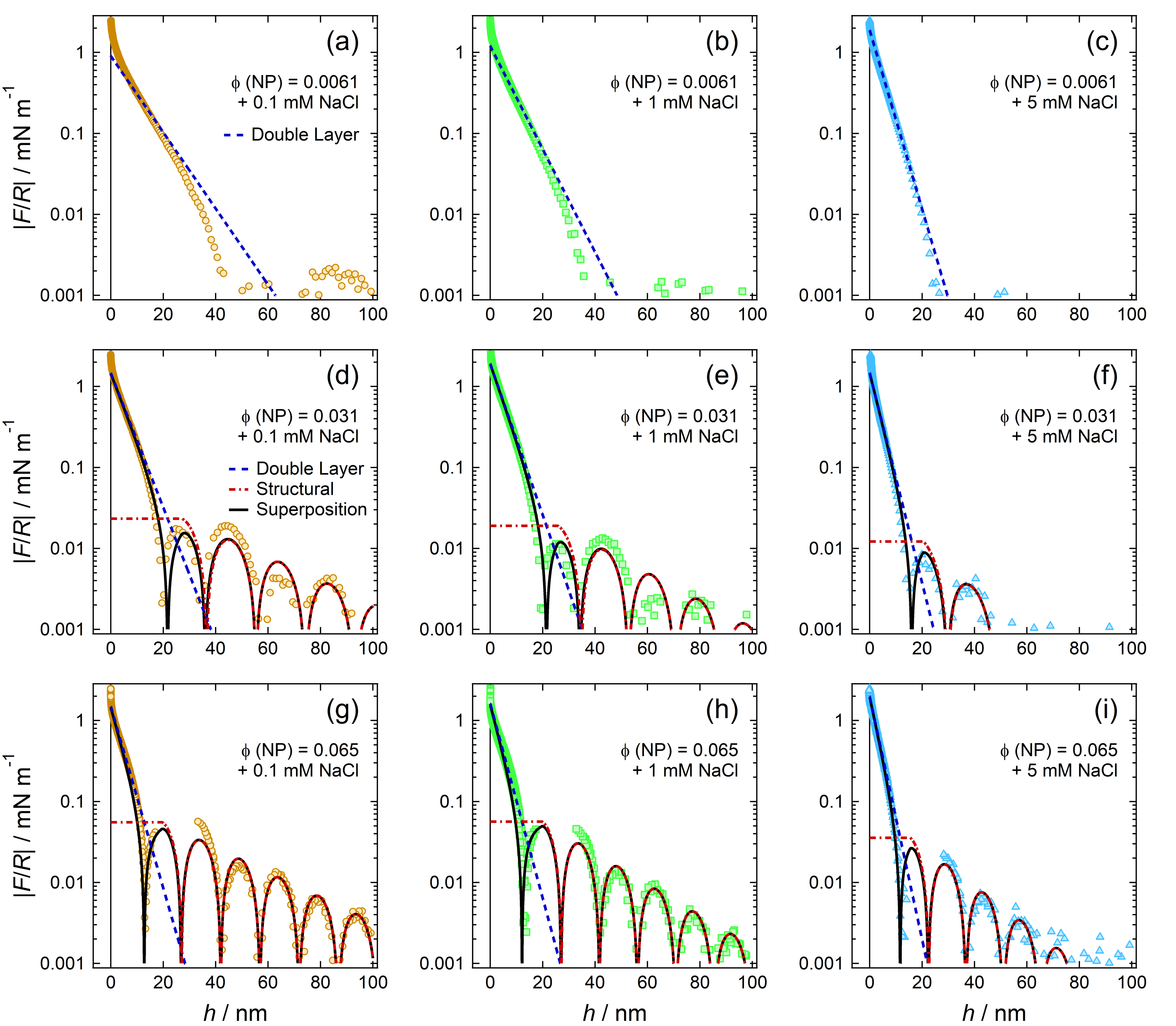}
\caption{Absolute values of the interaction forces between charged silica surfaces in aqueous suspensions containing like-charged silica NPs and salt at different concentrations. Experimental data (coloured dots) are modelled with a superposition (black, solid line) of double layer (blue, dashed line) and structural forces (red, dashed line). No structural forces could be fitted for the smallest volume concentration of silica NPs ($\phi$\,=\,0.0061). Double layer forces are fitted to equation \ref{equ:double_layer_force} by adjusting $\psi_{\text{eff}}$ as the only fit parameter.}
\label{fig:curves_log}
\end{center}
\end{figure}

Semilogarithmic representations of the absolute values of interaction forces between charged silica surfaces in aqueous suspensions containing like-charged silica NPs and salt are given in figure \ref{fig:curves_log}. The complete measured force up to the constant compliance region is visualized. This representation emphasises the behaviour of the double layer force as well as its transition region towards the structural force. 

The determination of the ionic strength in section \ref{sec:ionic_strength} allows the Debye screening length $\kappa^{-1}$ to be calculated independently with equation \ref{equ:debye_screening} which is put as a fixed parameter into the fit procedure. Double layer forces are fitted to equation \ref{equ:double_layer_force} by adjusting the effective surface potential $\psi_{\text{eff}}$.

Figure \ref{fig:curves_log}\,(a-c) shows the force profiles with NP volume fractions of $\phi$\,=\,0.0061. The force profile at 0.1 mM NaCl exhibits a deviation from the ideal, exponential decay, as known for pure NaCl solutions (figure S2, ESI). Upon further addition of NaCl this deviation decreases, until at 5 mM NaCl the interaction profile is well described by the exponential double layer force. Generally, the Debye screening length $\kappa^{-1}$ decreases with increasing salt concentration, resulting in shorter ranged double layer forces at higher concentrations of NaCl. 

Interaction profiles with NP volume fractions of $\phi$\,=\,0.031 are shown in Figure \ref{fig:curves_log}\,(d-f). Higher amounts of NPs involve increased amounts of counterions which result in a stronger Debye screening of the surface potential compared to smaller volume fractions of NPs. The Debye screening length of the double layer force decreases further upon addition of NaCl. Superposition of the double layer and structural forces results in a good description of the experimental data even in the transition region of both force contributions, \textit{i.e.} the region where both, the double layer and the structural force contribute considerably to the total interaction force.

Figure \ref{fig:curves_log}\,(g-i) shows the interaction profiles at NP volume fractions of $\phi$\,=\,0.065. The same trends in the double layer force are observed as compared to the ones with smaller NP volume fractions. The decay in the double layer forces are well described with the independently calculated ionic strength.

\begin{figure}[h!]
\begin{center}
\includegraphics[width=0.32\textwidth]{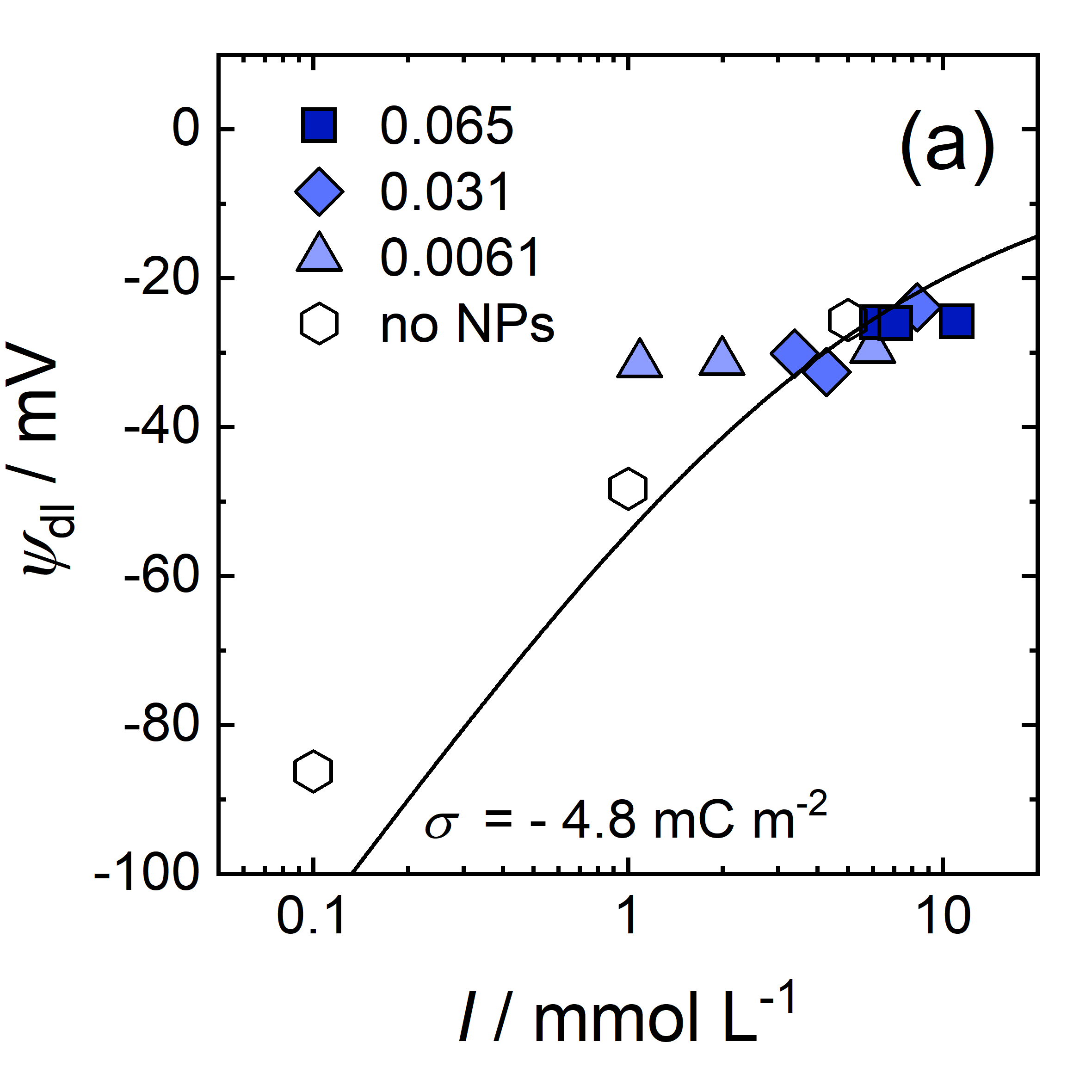}
\includegraphics[width=0.32\textwidth]{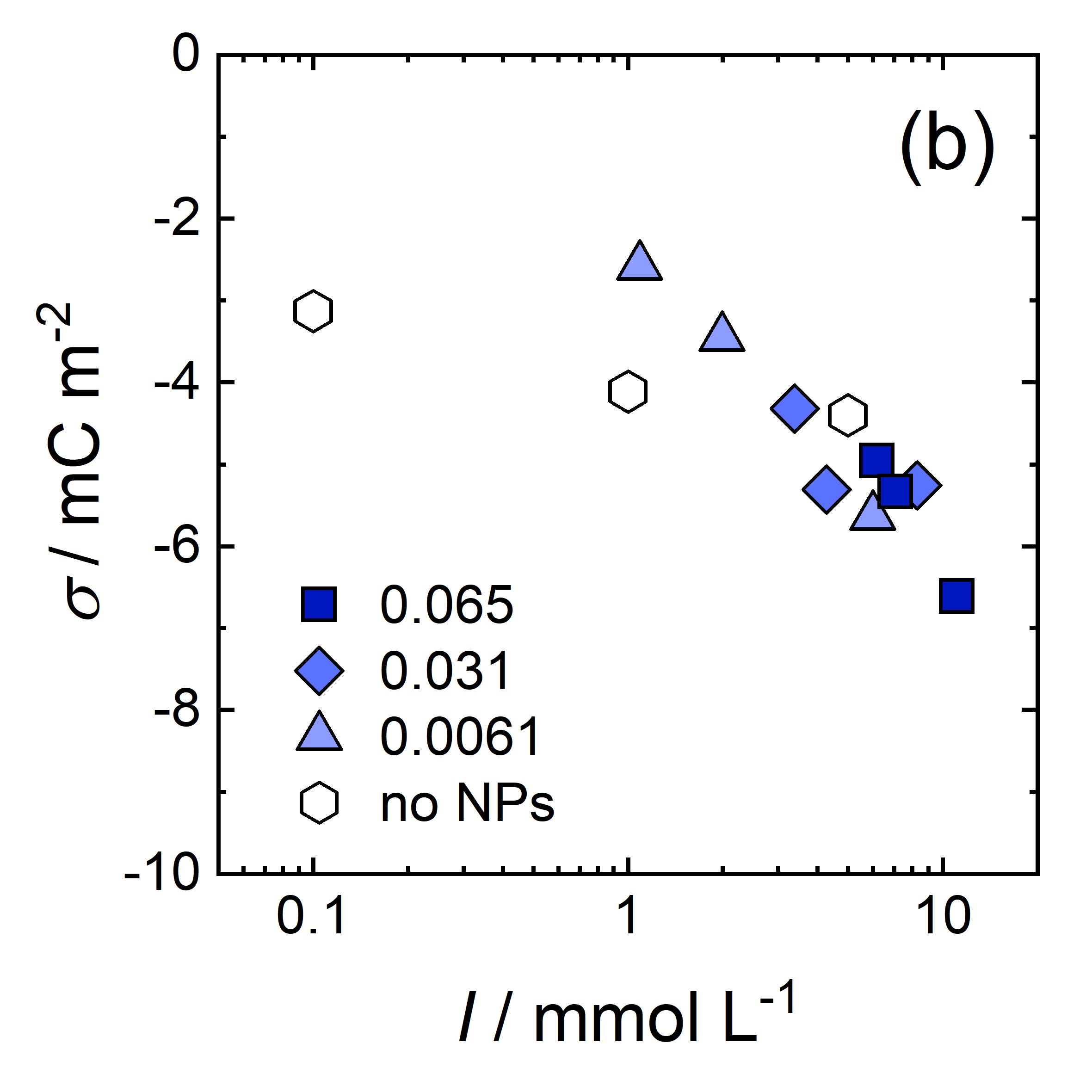}
\caption{Surface characteristics at different volume fractions of NPs and upon addition of NaCl. The total ionic strength $I$ is calculated as the sum of the ionic strengths of the pure NP suspension and added NaCl, $I = I_{0} + I_{\text{NaCl}}$. (a) The surface diffuse layer potential $\psi_{\text{dl}}$ as calculated from the fitted effective surface potential $\psi_{\text{eff}}$ via equation \ref{equ:diffuse_layer_pot}. (b) The surface charge density $\sigma$ as calculated from equation \ref{equ:Grahame}.}
\label{fig:fitparameter_dl}
\end{center}
\end{figure}

The total ionic strength $I$ is calculated according to equation \ref{equ:total_ionic_strength}, as the sum of the ionic strength of the pure NP suspension and the ionic strength of the added salt. The diffuse layer potential $\psi_{\text{dl}}$ is calculated from the fitted effective surface potential $\psi_{\text{eff}}$ via equation \ref{equ:diffuse_layer_pot}. Absolute values of the diffuse layer potential of the confining surfaces typically decreases with ionic strength in pure electrolyte solutions (data in figure S2, ESI). When NP are in solution, however, the effective surface potential appears rather constant with a mean value of $\psi_{\text{dl}}$\,=\,-\,28.6\,$\pm$\,2.9\,mV, irrespective of the concentration of NPs and salt in the system. It remains unclear if the addition of NPs constrain the effective potential of the confining surfaces to a constant value or if the effecive potential will change at different ionic strengths, since only a small interval of the ionic strength can be investigated.

Surface charge densities $\sigma$ of the confining surfaces were also calculated via the Grahame equation \ref{equ:Grahame}. Resulting values can be seen in figure \ref{fig:fitparameter_dl}\,(b). The absolute value of the surface charge density is found to increase with ionic strength. This behaviour was found before in pure electrolyte solutions \cite{Smith.2019b} and was attributed to \textit{e.g.} higher degree of ionization of surface groups, or dissociation of adsorbed water molecules. The mean value of the surface charge density in presence of NP suspensions is $\sigma$\,=\,-\,4.8\,$\pm$\,1.1\,mC\,m$^{-2}$. This is in good agreement with literature values \cite{MoazzamiGudarzi.2017,Smith.2019}. Under the assumption of this constant surface charge density, the general trend in the diffuse layer potential could be outlined, as visualized by the solid line in \ref{fig:fitparameter_dl}\,(a).

\subsubsection{Structural forces} \label{subsubsec:exp_structural}

\begin{figure}[p]
\begin{center}
\includegraphics[width=\linewidth]{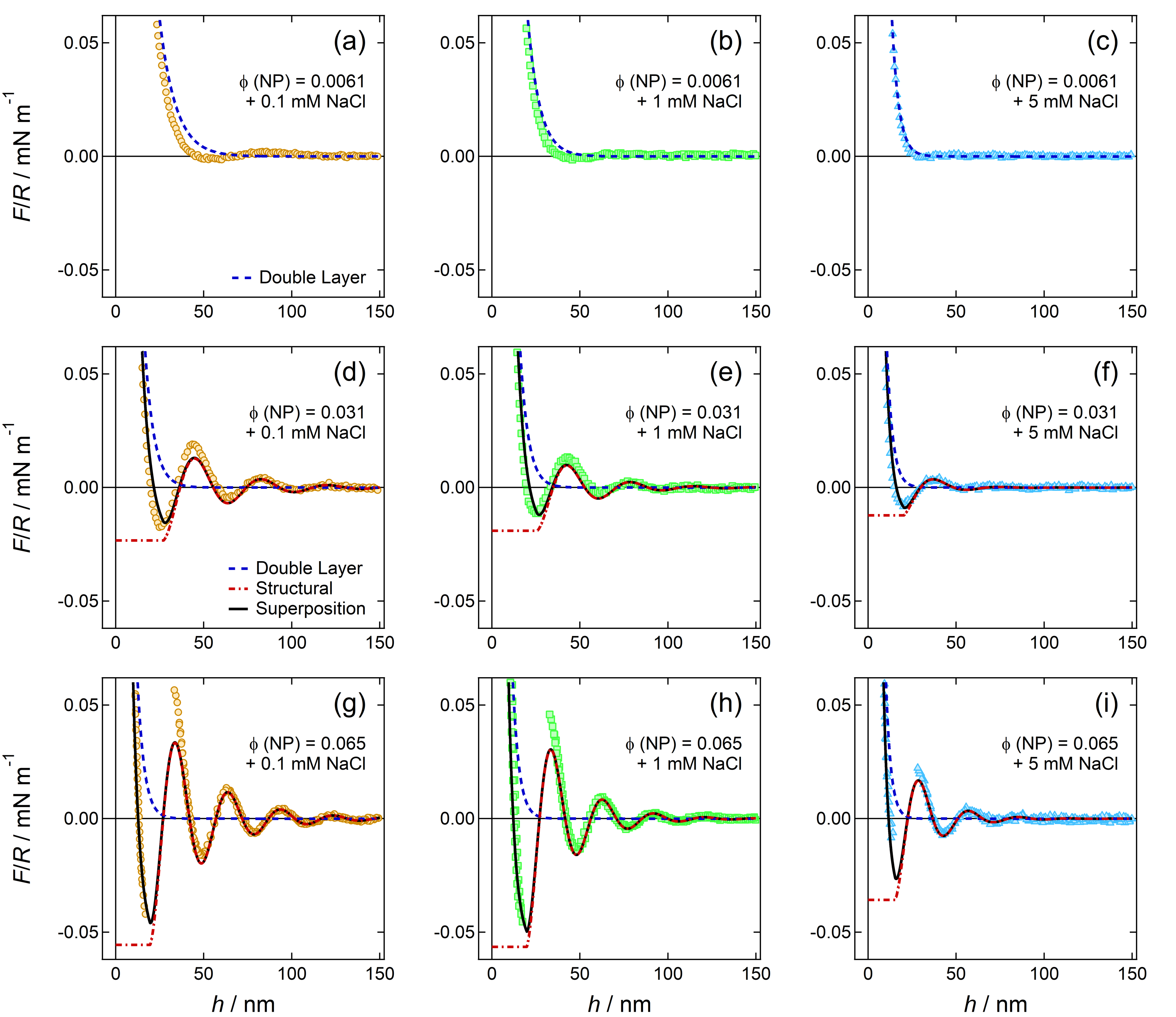}
\caption{Interaction forces between charged silica surfaces in aqueous suspensions containing silica NPs and salt at different concentrations, \textit{i.e.} the same force data as in figure \ref{fig:curves_log}. Linear representation increases the resolution of the structural force. Experimental data (coloured dots) is modelled with a superposition (black solid line) of double layer (blue, dashed line) and structural force (red, dashed line). No structural forces could be fitted for the smallest volume concentration of silica NPs ($\phi$\,=\,0.0061).}
\label{fig:curves_linear}
\end{center}
\end{figure}

The interaction force profiles in linear representation are shown in figure \ref{fig:curves_linear}. A smaller force range is chosen for a better resolution of the structural force.

Figure \ref{fig:curves_linear}\,(a-c) shows the force profiles at NP volume fractions of $\phi$\,=\,0.0061. At this concentration, the force profiles are almost completely determined by the double layer force. At 0.1\,mM added NaCl, a small contribution of the structural force can be observed. It is, however, too small to be reasonably described by a sufficient fit model in order to extract further information. Upon further addition of NaCl, the total ionic strength - from the NP's counterions and added salt - increases almost by a factor of six from (a) to (c). As a result, the structural force weakens, until at 5\,mM NaCl no structural force is observed and the interaction profile is well described by the double layer force only.

Interaction profiles at NP volume fractions of $\phi$\,=\,0.031 are displayed in figure \ref{fig:curves_linear}\,(d-f). With increased NP volume fraction, structural forces are more pronounced and fitting is possible using equation \ref{equ:structural_force} for all salt concentrations investigated. Three pronounced oscillations are observed in the structural force in the system with a low amount of added salt. With increasing salt concentration, the structural force contribution decreases, so that upon addition of 5\,mM only one clear oscillation remains observable. At the same time, the magnitude of the depletion attraction, \textit{i.e.} the force at surface separation below the first minimum in the force profile, decreases with increasing ionic strength in the systems.

Interaction profiles in suspensions of NP volume fractions of $\phi$\,=\,0.065 are shown in figure \ref{fig:curves_linear}\,(g-i). The structural force is most pronounced at the highest NP volume fraction. With more pronounced oscillations in the force profile, the gradient of the force exceeds the cantilevers spring constant between the first minimum and first maximum. As a result, the full force profiles cannot be resolved and a  jump-in (at approach) or jump-out (when retract) to the next stable force branch occurs between the first minimum and first maximum in the force profile. At least five force oscillations up to surface separations of more than 150\,nm are observed at low salt concentrations. Fitting of the structural force using equation \ref{equ:structural_force} systematically shows and underestimation of the first maximum in the force profile, which is most pronounced at highest volume fractions of silica and lowest salt concentration, \textit{i.e.} at most pronounced force oscillations.

\begin{figure}[ht]
\begin{center}
\includegraphics[width=0.32\textwidth]{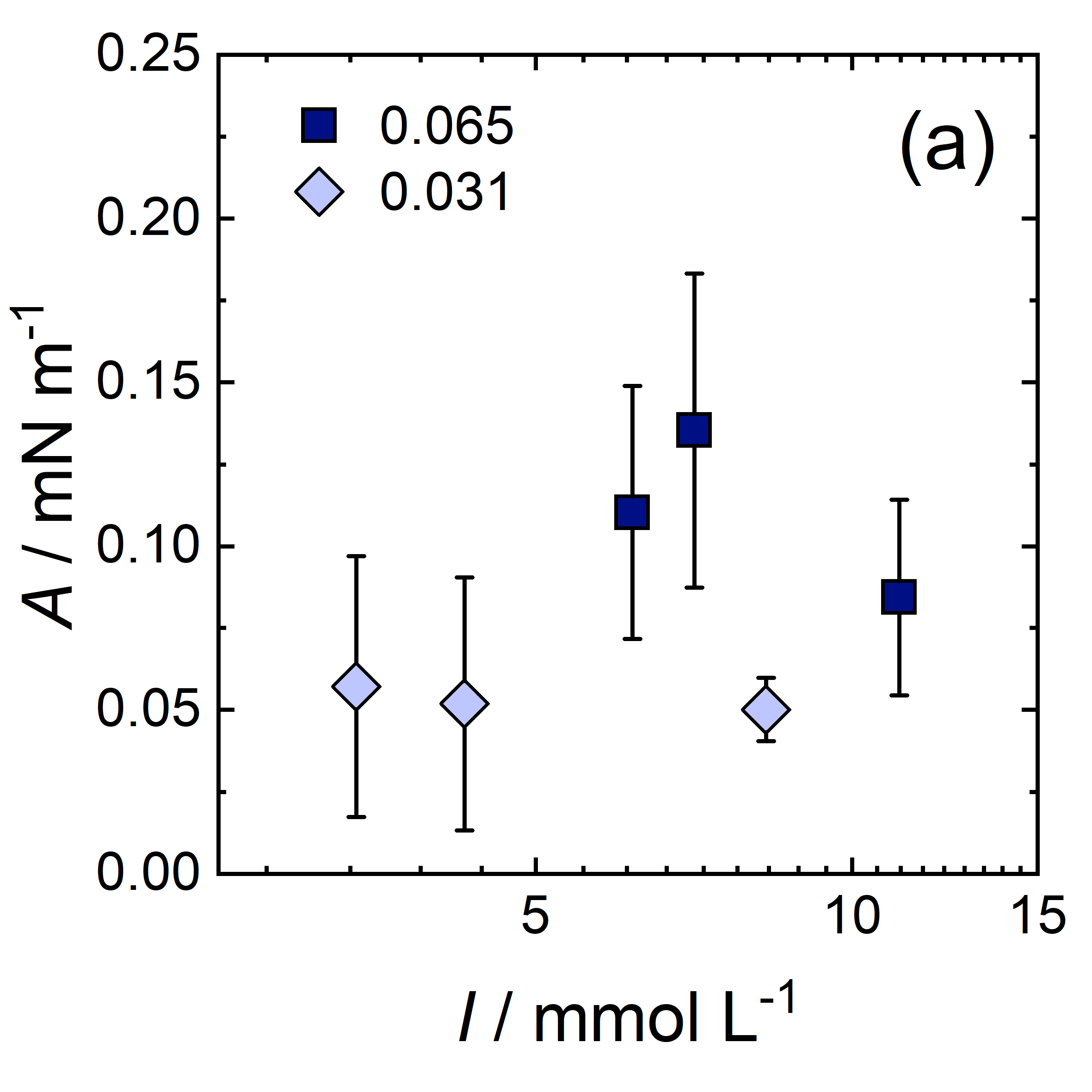}
\includegraphics[width=0.32\textwidth]{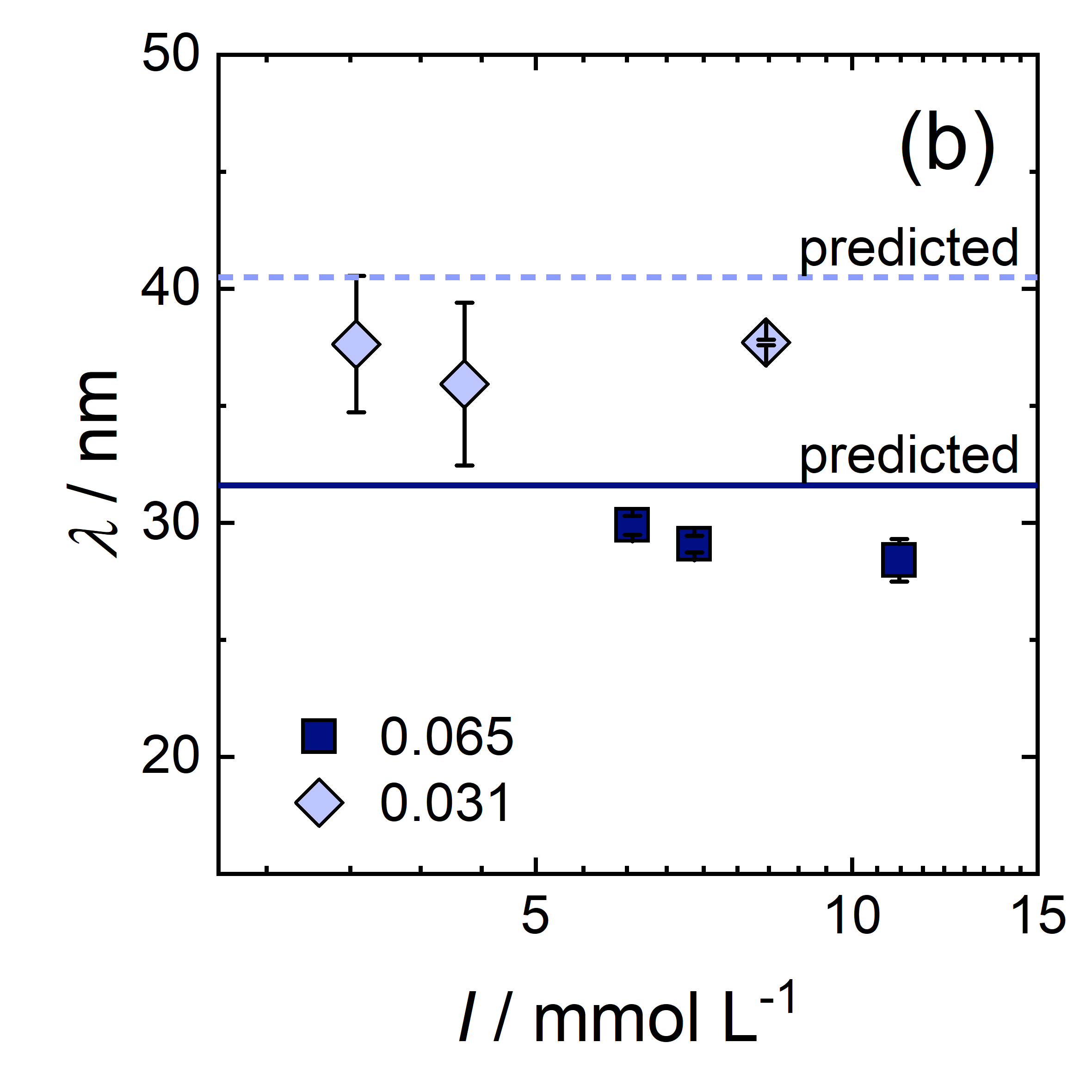}\\
\includegraphics[width=0.32\textwidth]{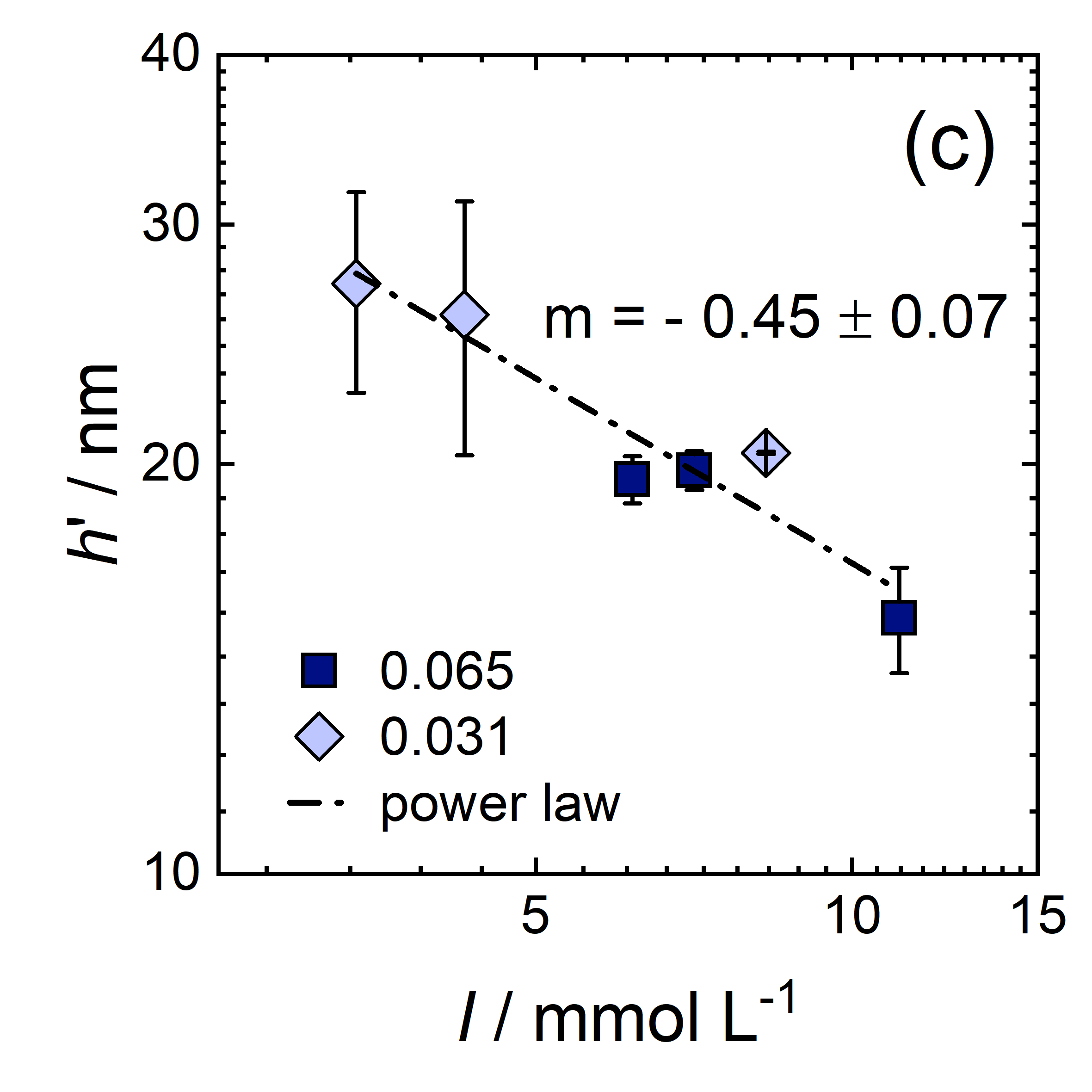}
\includegraphics[width=0.32\textwidth]{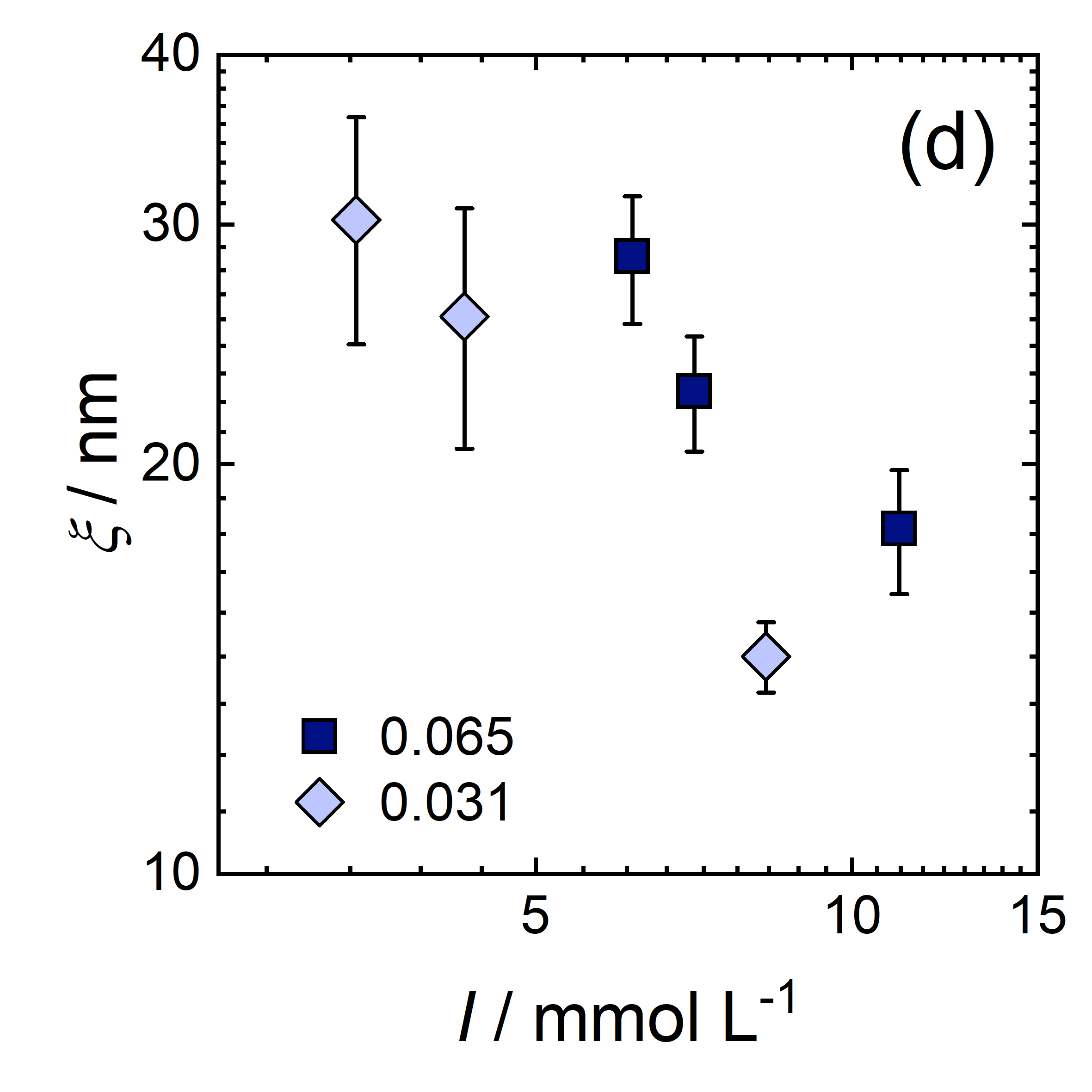}
\caption{Fit parameters of the structural force at different volume fractions of NPs and upon addition of NaCl. Total ionic strength is calculated as the sum of added NaCl and the ionic strength of the pure NP suspension. Parameters describing the structural forces are (a) amplitude $A$, (b) wavelength $\lambda$, (c) offset $h'$ and (d) the decay length $\xi$. In (b), the wavelength $\lambda$ can also be predicted using the inverse cubic root scaling law $\lambda$\,=\,$\rho^{-1/3}$ (dashed line: $\phi$\,=\,0.031, solid line: $\phi$\,=\,0.065). Fitting was performed in (c) to extract a scaling exponent m for $h'\,\propto\,I^{\text{m}}$.}
\label{fig:fitparameter_dep}
\end{center}
\end{figure}

The fit parameters for the structural forces are given in figure \ref{fig:fitparameter_dep}. The amplitude $A$ is known to increase with higher volume fractions of silica \cite{Zeng.2011}. This finding can be confirmed in this study. No distinct change in amplitude is obtained upon addition of monovalent salt. The wavelength $\lambda$ is known to be related to the particle number density $\rho$ of NPs in the suspension via the inverse cubic root scaling law: $\lambda$\,=\,$\rho^{-1/3}$ \cite{Tulpar.2006,Ludwig.2019} and is independent of the ionic strength in the suspension. \cite{Klapp.2010} The number particle density can independently be calculated from the volume fraction and diameter of the particles. The results are displayed as dashed lines in figure \ref{fig:fitparameter_dep}\,(b). Fitted values are a few percent below the predicted values. The offset $h'$ generally decreases with increasing ionic strength in suspension. A scaling exponent m\,=\,-\,0.45\,$\pm$\,0.07 for $h'\propto\,I^{\text{m}}$ could be extracted and will be further considered in the discussion section. The decay length $\xi$ decreases with inceasing ionic strength. A higher volume fraction and, consequently, a tighter packing of the particles increases the decay length. This in good agreement with simulations from Klapp \textit{et al.} \cite{Klapp.2008}.

\subsubsection{Superposition of double layer and structural force}

The complete force profile can be divided into different parts, according to the prevalent force contributions. At small surface separations, the interaction force is determined by the strength of the double layer force with its typical exponential decay when the Debye H\"uckel (DH) approximation is applied. The interaction force at larger surface separations, however, is purely determined by the structural force, since it is typically longer-ranged than the double layer force.

In between these parts a transition region is observed, where both, the double layer and the structural force contribute considerably to the total interaction force. Some features may seem surprising when looking at it from a perspective of a single force contribution. On the one hand, deviations from an exponential decay at smaller surface separation are observed once the structural force gains influence compared to the double layer force. While on the other hand, the oscillations in the force profile are asymmetric to the zero-force line in the region where still a considerable strength of the double layer force is present. Those features are no properties of the force contributions itself but originate from the superposition of both forces.

\section{Discussion}

\subsection{Model of the double layer force using jellium approximation}

The challenge of describing the screening behaviour of suspensions containing charged nanoparticles (NPs) and salt is that the system represents a strongly asymmetric electrolyte of small monovalent ions and large multivalent particles (macroions). Different approaches to this problem have already been published. Danov \textit{et al.} compared the applicability of the Poisson-Boltzmann model (PB) with the jellium approximation (JA) for the description of the screening behaviour in dispersions containing ionic micelles \cite{Danov.2011}. For the general discussion here, the framework of dispersions containing macroions and added salt is used.

In the PB model, all charge carriers in the dispersion (monovalent ions and multivalent macroions) obtain Boltzmann distribution as a response to the surfaces electric field. As a result, both the multivalent macroions and the monovalent ions contribute to the Debye screening. The valence per particles $z_{i}$ can be very large, resulting in a very strong Debye screening when applying equation \ref{equ:ionic_strength}.

Moazzami-Gudarzi \textit{et al.} described the interaction profile in dispersions containing strong, negatively charged polyelectrolytes (macroions) and added monovalent salt with a superposition of double layer and structural forces \cite{MoazzamiGudarzi.2016, MoazzamiGudarzi.2017}. A numerical solution of the PB model was chosen for the description of the double layer interaction. Two distinct regions are defined: (i) At small surface separations, the polyelectrolytes are too large to enter the vicinity between the surfaces and Debye screening is only transacted through the monovalent ions in the dispersion. (ii) Once the surface separation is large enough that the macroions can enter the vicinity between the confining surfaces, they also contribute to Debye screening and the double layer force typically vanishes within a few nanometer. The non-exponential decay in the force profile was described as the non-exponential double layer force. The effective charge per macroion enters the fit of the double layer force as a free parameter. Using this description, almost no transition region between the double layer force and structural force dominated regions exist.

In the PB model, monovalent ions and multivalent macroions are treated as point-like ions. In reality, the macroions themselves have a considerable volume. Moreover, the repulsive interactions between the macroions hinder them from a Boltzmann distribution. Danov \textit{et al.} concluded, that the JA offers an improved description of the screening behaviour compared to the PB model. In this approximation, the macroions obtain a uniform distribution and the macroions are excluded from Debye screening. Only the co- and counterions of low valency contribute to it. JA was already used - without specifically calling it JA - to calculate the interparticle interactions in dispersions containing polyelectrolytes \cite{Klitzing.1999} and nanoparticles \cite{Klapp.2008b}, yet without description of the interaction of the confining surfaces.

The non-exponential decay of the total interaction force in this study (see figure \ref{fig:curves_log}) is caused by the superposition of the exponentially decaying double layer force and the structural forces. When the macroions are excluded from Debye screening, DH approximation shows reasonable results, with a reduction of fit parameters based on an independent determination of the ionic strength of the dispersion. This result offers an easy description of the double layer force when its superposed with structural forces, together with the interesting finding that the macroions themselves \textit{do not} contribute to Debye screening. However, macroions do contain a considerable amount of counterions (in this study up to a concentration of a few millimolar), which \textit{do} contribute to Debye screening. It is, therefore, irrelevant for the double layer force if the macroions are present in confinement.

\subsection{Model of the structural force and the influence of a particle-free layer}

Maroni \textit{et al.} recently studied the structuring of the same type of silica NP suspensions (Ludox HS) near the suspension-silica interface, using specular neutron reflectivity \cite{Maroni.2020}. The concentration profile of the NPs perpendicular to the interface is characterised by a damped oscillatory profile, indicating a layering of the particles next to a wall. In order to correctly fit the reflection curves, a particle free layer next to the interface was defined. The thickness of this particle free layer $d$ depends on the volume fraction of the dispersed silica particles $\phi$ with a scaling law of roughly $d\,\propto\,\phi^{-1/2}$. The authors concluded that the dependency originated from the electrostatic repulsion of the NPs from the charged surface.

\begin{figure}[ht]
\begin{center}
\includegraphics[width=0.4\textwidth]{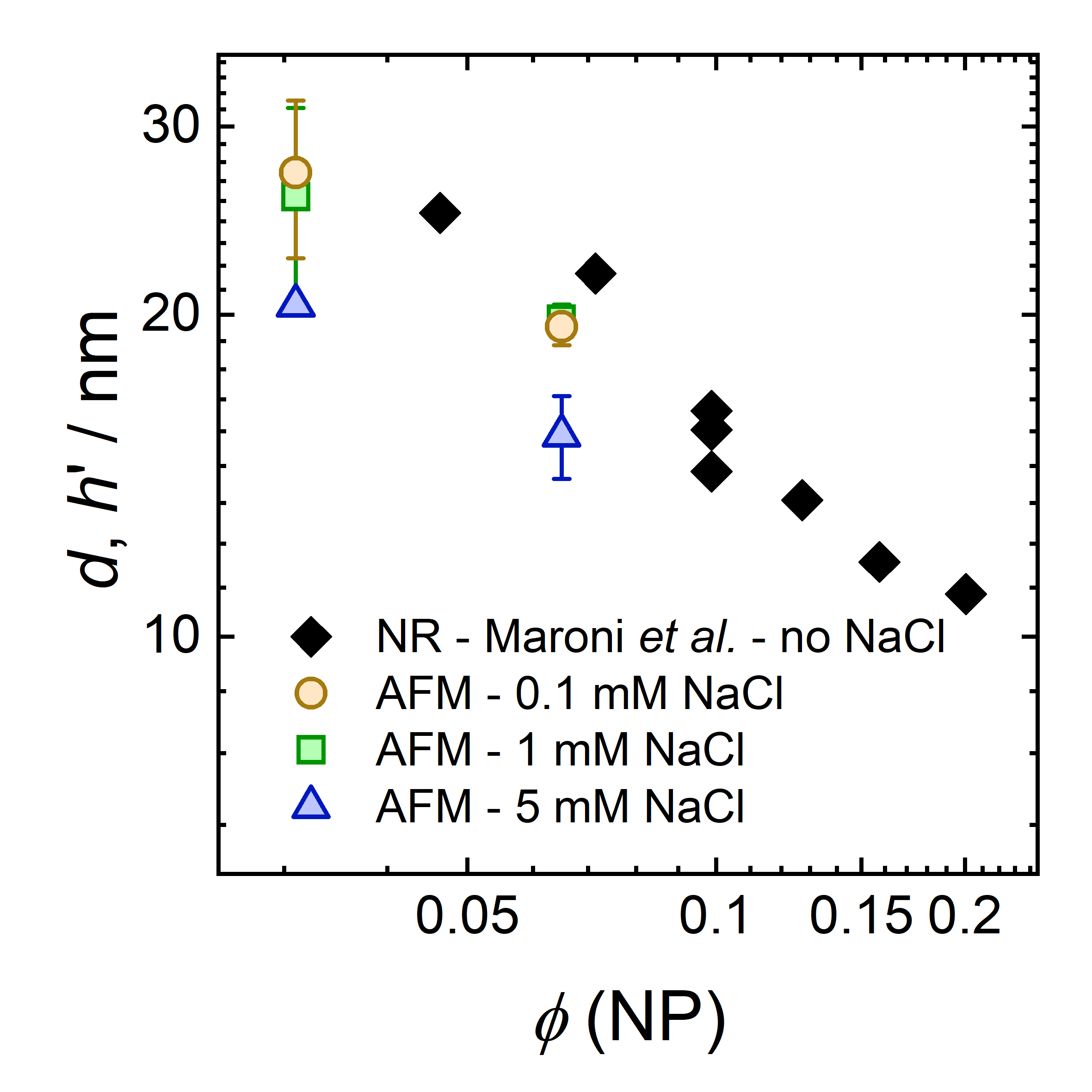}
\caption{Comparison of the thickness $d$ of particle free layers next to a single silica-liquid interface as probed by specular neutron reflectivity from Maroni \textit{et al.} \cite{Maroni.2020} (black dots, without NaCl) with the offset $h'$ extracted from fitting the structural force (coloured dots, at various salt concentrations). Values are presented as a function of the volume fraction $\phi$ of NPs in suspension.}
\label{fig:particle_free_layer}
\end{center}
\end{figure}

The oscillatory structural force is the direct result of overlapping concentration profiles, when two surfaces are brought into close contact. Therefore, a particle free layer at one surface should be directly related to an offset in the oscillatory structural force between two surfaces, as described in this study. 

Figure \ref{fig:particle_free_layer} shows the comparison of the particle free layer thickness $d$ (from neutron reflectivity data) with the offset $h'$ extracted from the position of the first minimum of force measurements. Both experiments are in good agreement with each other. With increasing amount of added salt, the electrostatic screening increases due to an increased ionic strength. This results in a decrease in the thickness of the particle free layer and, therefore, also in the offset (see also figure \ref{fig:fitparameter_dep}\,(c)). The scaling exponent m\,=\,-\,0.45\,$\pm$\,0.07 for $h'\propto\,I^{\text{m}}$ confirms the finding of Maroni \textit{et al.} for the offset being of electrostatic origin. A scaling exponent of -\,0.5 is rationalized by the Debye screening length, which increases with the inverse of the square root of the ionic strength. The offset can also be compressed by addition of salt at a constant volume fraction of NPs. This further indicates that the offset is rather dominated by the electrostatic interactions between the particles and the surfaces, than by simple packing effects of the particles in the fluid.

The determination of a particle free layer next to the surface also affirms the introduction of the depletion attraction once the surface separation is smaller than the offset, since all particles are considered depleted from the vicinity of the confining surfaces.

\subsection{Deviation between fit and data}

With increasing NP concentration and decreasing ionic strength, deviations of the fit and the experimental data increase. The deviation is, therefore, most pronounced in figure \ref{fig:curves_linear}\,(g). An underestimation of the data by the fit often occurs at the last layer of NPs, especially, when they are well ordered, \textit{i.e.} at high concentration and high repulsion between the NPs. Similar underestimations have also been reported for highly compressed ionic liquids \cite{Hoth.2014}.

Although an extended fitting formula that accounts for an additional repulsive contribution in the oscillatory structural force was presented recently \cite{Schon.2018,Schon.2018b}, it is excluded for this analysis in order to keep the fit model as simple as possible. The underestimation of the first maximum is considered as the breakdown of the asymptotic behaviour, predicted by density functional theory (DFT) for large distances \cite{Roth.2000}. This deviation cannot be explained by purely electrostatic reasons and might be seen as an confinement effect itself, which has to be further clarified.

\section{Conclusion}

Interaction forces between negatively charged surfaces across suspensions containing nanoparticles and monovalent salt are successfully measured using colloidal-probe atomic force microscopy. Full description of the force profiles is achieved by the superposition of double layer and structural forces with no mutual effect between the two types of forces. Following conclusions are made:

The charged nanoparticles show a preferred self-organized structure which can be described by the jellium approximation. Next to a surface, they form a layer-like arrangement which can be detected as oscillatory structural force. The particles are depleted from the region next to a confining surface. The thickness of this depletion zone is dominated by the electrostatic repulsion of the particles from the like-charged surface rather than by particle concentration, \textit{i.e.} packing effects. This picture is in good agreement with a particle-free layer next to a single surface as probed by neutron reflectivity measurements. Unlike the nanoparticles, the monovalent ions (the particles counterions and added salt ions) follow Boltzmann distribution. Debye-H\"uckel approximation was sufficient to describe the double layer force.

It has to be emphasized that the total force profile can be simply divided into a pure double layer force and a pure structural force contribution. The decay of the double layer force is only determined by the Boltzmann distributed monovalent ions while the structural force originates simply from the layer-like arrangement of the particles. This means that the different distributions of particles and ions do not affect each other. Deviations from an exponential decay at small separations, as already reported in literature, are simply caused by the superposition of a Debye-H\"uckel double layer and the structural force. The same superposition leads to asymmetry of the force oscillation.

The study helps understanding surface interactions across different kinds of colloidal dispersions as superposition of DLVO and structural forces. The question rises if this simple model describes also other types of confined colloidal dispersions like for instance biomolecules in a physiological solution between biomembranes or complex fluids in wetting films.

\section*{Conflicts of interest}
There are no conflicts to declare.

\newpage
\section*{Supporting Information (ESI)}
\setcounter{figure}{0}
\setcounter{table}{0}
\setcounter{equation}{0}
\makeatletter 
\renewcommand{\thefigure}{S\@arabic\c@figure}
\renewcommand{\thetable}{S\@arabic\c@table}
\renewcommand{\theequation}{S\@arabic\c@equation}
\makeatother

\paragraph*{Potentiometric titration of nanoparticle suspensions}

Potentiometric titration of pure nanoparticle (NP) suspensions (initial volume = 10\,mL) at three different volume concentrations of NPs was carried out upon addition of hydrochloric acid solution (HCl, Titrisol, Merck, Germany). The pH was measured using a pH-electrode (inoLab pH 720, VWR).

\begin{figure}
\begin{center}
\includegraphics[width=0.4\linewidth]{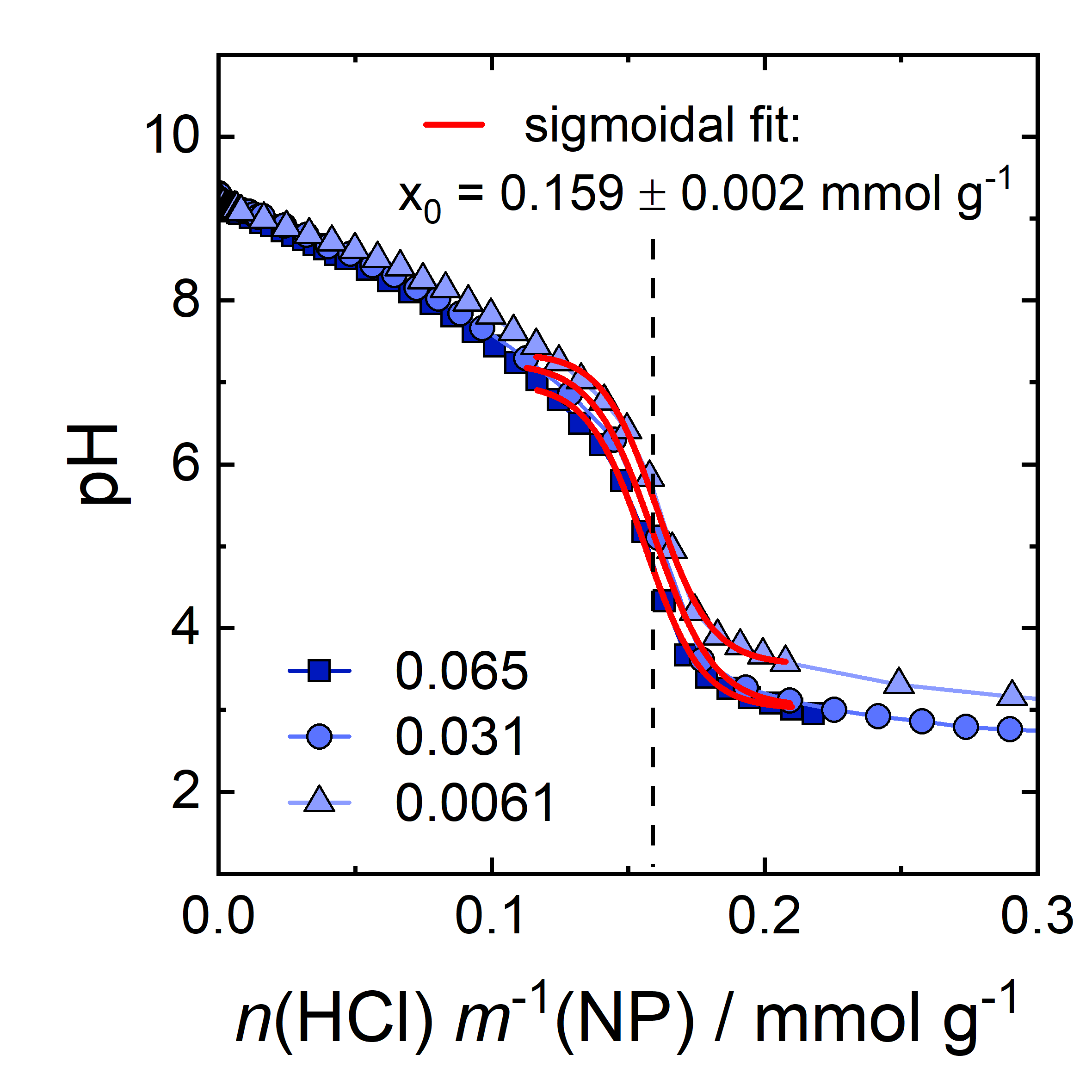}
\caption{Potentiometric titration of pure NP suspensions (without added salt) upon addition of HCl for various volume fractions of NPs. The amount of added HCl is normalised by the total mass of NPs in the suspension. Sigmoidal fitting is used to extract the point of inflection $x_{0}$ with the mean value being displayed in the graph.}
\label{fig:titration}
\end{center}
\end{figure}

The surface charge density $\sigma$ of NPs in pure, aqueous dispersions can be calculated from the point of inflection $x_0$ in the titration curve (0.159\,$\pm$\,0.002\,mmol\,g$^{-1}$), the density of silica NPs $\rho$ and the mean diameter $d$ of the NPs, assuming spherical particles.

\begin{equation} \label{equ:surf_charge_calc}
\sigma = x_{0} N_{\text{A}} \rho \, \frac{d}{6}
\end{equation}

Equation \ref{equ:surf_charge_calc} results in a surface charge density $\sigma = 0.50\,\pm\,0.01\,\text{nm}^{-2} = 0.080\,\pm\,0.002\,\text{C\,m}^{-2}$, irrespective of the NP concentration. This is in good agreement with literature values \cite{Bolt.1957,Dove.2005,Brown.2016b}.

\paragraph*{Interaction forces between charged surfaces in pure electrolyte solution}

Interaction forces of negatively charged silica surfaces in pure NaCl solutions were measured for reference.

\begin{figure}
\begin{center}
\includegraphics[width=\linewidth]{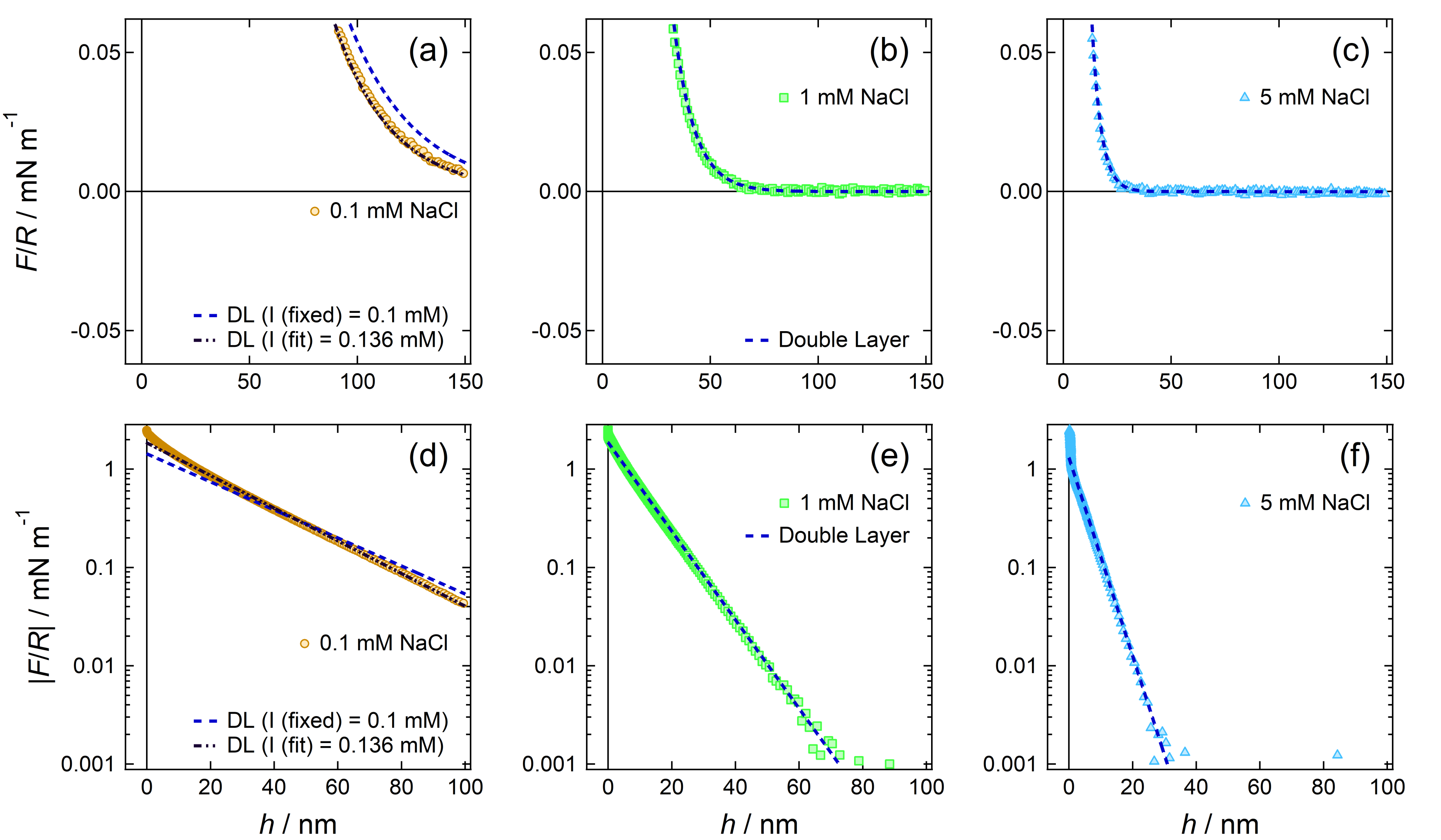}
\caption{Interaction forces between silica surfaces in aqueous NaCl solutions. Experimental data (dots) is compared with the fitted double layer force (dashed line). (a-c): linear representation, (d-f): semilogarithmic representation for emphasis of the double layer force. For 0.1\,mM NaCl (a,d) two different fits were performed with either the ionic strength $I$ being fixed or entering as free fit parameter of the double layer force.}
\label{fig:force_pure_NaCl}
\end{center}
\end{figure}

Figure \ref{fig:force_pure_NaCl} shows the measured interaction force profiles in pure NaCl solutions. While (a-c) is in linear representation, (d-f) shows the same data in semilogarithmic representation, for emphasis of the double layer force. The blue dotted lines are the fit to the double layer force, equation \ref{equ:double_layer_force}, with the effective surface potential $\psi_{\text{eff}}$ being the only parameter to enter the fitting. Effective surface potentials $\psi_{\text{eff}}$ are determined as -\,70.0\,$\pm$\,4.4\,mV, -\,44.9\,$\pm$\,1.5\,mV, and -\,25.1\,$\pm$\,0.9\,mV in 0.1\,mM, 1\,mM, and 5\,mM NaCl solutions, respectively. This is in good aggreement with literature values \cite{Smith.2019b}. In figures \ref{fig:force_pure_NaCl}\,(a,d) also a black dotted line is shown, which represents the fit of the double layer force with the ionic strength $I$ also as free fit parameter. Best fit result is obtained for an ionic strength $I$ of 0.136\,mM which shows the difficulty to work at such low ion concentrations, since \textit{e.g.} adsorption of \ch{CO2} may change the ionic strength to this amount.

\newpage
\bibliography{references}

\end{document}